\documentclass[preprint2]{aastex701}
\usepackage{multirow}
\usepackage{soul}
\usepackage{CJK}
\usepackage{enumitem}
\usepackage{graphicx}
\usepackage{subcaption}
\usepackage{siunitx}
\usepackage{amsmath}
\DeclareSIUnit\au{au}
\usepackage{xspace}
\newcommand{\kamo}{Kamo\textquoteleft{}oalewa\xspace}
\newcommand{\tnm}[1]{\,\tablenotemark{#1}}
\newcommand{\tntext}[2]{\tablenotetext{#1}{\ \ #2}}

\begin{document}
\begin{CJK*}{UTF8}{gbsn}

\title{The Non-Principal-Axis Rotation and Convex Shape Model of Earth Quasi-Satellite and the Target of China's Tianwen-2 Mission (469219) \kamo}

\shorttitle{Non-Principal-Axis Rotation of \kamo}

\author[sname=Sun, gname=Xiaoyu]{Xiaoyu Sun (孙效宇)}
\affiliation{School of Aeronautic Science and Engineering, Beihang University, Beijing, People's Republic of China}
\email{sunxiaoy@buaa.edu.cn}

\author[gname=Zhijun, sname=Song]{Zhijun Song (宋志军)} 
\affiliation{School of Aeronautic Science and Engineering, Beihang University, Beijing, People's Republic of China}
\email{junzhisong@buaa.edu.cn}

\author[orcid=0000-0002-6144-3062, gname=Hanjie, sname=Tan]{Hanjie Tan (谭瀚杰)}
\affiliation{Planetary Environmental and Astrobiological Research Laboratory (PEARL), School of Atmospheric Sciences, Sun Yat-sen University, Zhuhai, Guangdong, People's Republic of China}
\email{h.tan1996@hotmail.com}

\author[orcid=0000-0002-5033-9593, sname=Yang, gname=Bin]{Bin Yang (杨彬)}
\affiliation{Instituto de Estudios Astrof\'{i}sicos, Facultad de Ingenier\'{i}a y Ciencias, Universidad Diego Portales, Santiago, Chile}
\affiliation{Planetary Science Institute, 1700 East Fort Lowell, Suite 106 Tucson, AZ, USA}
\email{bin.yang@mail.udp.cl}

\author[orcid=0000-0003-4914-3646, gname=Josef, sname=\v{D}urech]{Josef \v{D}urech}
\affiliation{Charles University, Faculty of Mathematics and Physics, Institute of Astronomy, V Hole\v{s}ovi\v{c}k\'{a}ch 2, 180 00 Prague, Czech
Republic}
\email{durech@sirrah.troja.mff.cuni.cz}

\author[orcid=0000-0001-6765-6336, gname=Nicholas, sname=Moskovitz]{Nicholas Moskovitz}
\affiliation{Lowell Observatory, 1400 West Mars Hill Road, Flagstaff, AZ, USA}
\email{nmosko@lowell.edu}

\author[orcid=0000-0002-1506-4248, sname=Thirouin, gname=Audrey]{Audrey Thirouin}
\affiliation{Lowell Observatory, 1400 West Mars Hill Road, Flagstaff, AZ, USA}
\email{thirouin@lowell.edu}

\author[orcid=0009-0006-1160-3829, sname=Hemmelgarn, gname=Samantha]{Samantha Hemmelgarn}
\affiliation{Lowell Observatory, 1400 West Mars Hill Road, Flagstaff, AZ, USA}
\affiliation{Northern Arizona University, Flagstaff, AZ, USA}
\email{shemmelgarn@lowell.edu}

\author[gname=Wen, sname=Bo]{Wen Bo (博文)}
\affiliation{Planetary Environmental and Astrobiological Research Laboratory (PEARL), School of Atmospheric Sciences, Sun Yat-sen University, Zhuhai, Guangdong, People's Republic of China}
\email{bowen2@uchicago.edu}

\author[orcid=0000-0001-9329-7015, gname=Yu, sname=Yang]{Yang Yu (于洋)}
\affiliation{School of Aeronautic Science and Engineering, Beihang University, Beijing, People's Republic of China}
\email[show]{yuyang.thu@gmail.com}

\author[orcid=0000-0003-3841-9977,gname=Jian-Yang, sname=Li]{Jian-Yang Li (李荐扬)}
\affiliation{Planetary Environmental and Astrobiological Research Laboratory (PEARL), School of Atmospheric Sciences, Sun Yat-sen University, Zhuhai, Guangdong, People's Republic of China}
\affiliation{Xinjiang Astronomical Observatory, Chinese Academy of Sciences, Urumqi, People's Republic of China}
\email[show]{lijianyang@mail.sysu.edu.cn}

\shortauthors{Sun et al.}
\correspondingauthor{Yang Yu, Jian-Yang Li}


\begin{abstract}

(469219) \kamo is the most stable Earth quasi-satellite and the target of China's Tianwen-2 asteroid sample return mission.  Due to its small size, fast rotation, and the limited observing geometry accessible from the ground, many physical properties of \kamo remain poorly constrained, including the rotational state and shape.  We obtained three epochs of high-cadence, high signal-to-noise photometric lightcurves of \kamo with the Gemini North Telescope from 2026 April to May, supplemented by one lightcurve from the Lowell Discovery Telescope in 2026 May.  Our analysis suggests that \kamo is in a non-principal-axis rotation with an elongated shape.  Four possible solutions exist, including a long-axis mode (LAM) solution and a short-axis mode (SAM) solution, as well as their corresponding mirrored angular momentum directions.  The most preferable solution has a LAM model with a precession period $P_\phi=\qty{27.65(0.03)}{\min}$ and a rotational period $P_\psi=\qty{50.49(0.08)}{\min}$, and the angular momentum points to ecliptic coordinates \((\lambda, \beta) = (\qty{226}{\degree} \pm \qty{20}{\degree}, \qty{-39}{\degree} \pm \qty{20}{\degree})\), although we cannot rule out other solutions or other close-by periods due to aliasing.  We also derived a convex shape inversion for the LAM models with consistent rotational parameters but could not find a satisfactory inversion for the SAM models.  The corresponding angular momentum points to \((\lambda, \beta) = (\qty{225}{\degree}, \qty{-43}{\degree})\), and the periods are $P_\phi=\qty{27.90}{\min}$ and $P_\psi=\qty{49.66}{\min}$.  The non-principal-axis rotation provides additional constraints on the dynamic history or the internal structure of \kamo.

\end{abstract}



\section{Introduction}

Near-Earth asteroid (469219) \kamo represents the most dynamically stable member of the Earth quasi-satellite population \citep{2016MNRAS.462.3441D, 2021AJ....162..227F}.  Seven Earth quasi-satellites have been identified to date \citep{2025RNAAS...9..235D}.  They belong to the rare population of Earth co-orbital asteroids (ECAs) that share Earth's orbit in a 1:1 resonance \citep{2013CeMDA.117...91M}.  The ECAs provide a unique laboratory for probing the dynamics and material exchange of the inner Solar System, offering clues to the history of the Solar System and the delivery of water and organics to Earth.  They are also among the most accessible targets for scientific exploration and resource utilization.

\kamo is of particular interest because it is the target of China's Tianwen-2 asteroid sample return mission, which just arrived in early July 2026 and has started its global survey before sampling operations and sample return in 2027 \citep{2025SSPMA..55A9501Z, 2026SSRv..222...11Z}.  Highly limited knowledge of \kamo has been obtained due to the difficulties of observing it from the ground because of its faintness.  The only available ground-based spectral data \citep{2021ComEE...2..231S} have triggered the debate about whether it originates from the Moon or the main asteroid belt \citep[e.g.,][etc.]{2021ComEE...2..231S, 2024NatAs...8..819J, 2026Innov...701183Z, 2026ApJ...997L..49L, zhang_tianwen-2_2026}.  Recent observations from the James Webb Space Telescope (JWST) suggest a possible E- or V-type composition with a higher albedo ($p_V\sim0.6$) and smaller size (mean diameter $D\sim\qty{18}{\m}$) of \kamo \citep{2026arXiv260624017S} than previous estimates, adding further mysteries.

The shape and rotational state of an asteroid contain unique information on its dynamical evolution and collisional history.  They are important conditions to determine the evolution and distribution of the regolith on its surface \citep[e.g.,][etc.]{2015aste.book..509V, 2015aste.book..745S}.  The rotational state of \kamo also plays a critical role in the mission operations and sampling operations of Tianwen-2.  However, such information also remains largely obscured.  Reliable inversion of the shape and rotational state of an asteroid from lightcurves generally requires a sufficiently wide range of viewing and illumination geometries \citep[c.f.,][]{2015aste.book..183D}.  However, the Earth quasi-satellite nature of \kamo means that the observing aspects in every apparition remain similar.  Therefore, although a more or less consistent lightcurve period of $\sim\qty{28}{\min}$ has been determined \citep{2021ComEE...2..231S, 2026A&A...710L..35B, 2026arXiv260624017S}, the solutions to its spin pole orientation and shape have hardly converged \citep[e.g.,][]{2026A&A...710L..35B, 2025A&C....5100925Z, 2021Icar..35714249L, 2026arXiv260625512W}.  We note that all previous work assumed a principal-axis rotation with a fixed pole orientation, because the available datasets did not provide sufficient evidence to require a more complex rotational model.

Here we report the lightcurve observations of \kamo from the ground using the Gemini North Telescope (GN), supplemented by the Lowell Discovery Telescope (LDT) observations, conducted from April to May 2026.  The large mirror of GN allows us to obtain lightcurve data at a high cadence of \qtyrange{1}{2}{\min} with signal-to-noise ratios (SNRs) of \numrange{10}{15}.  Our analysis suggests that a non-principal-axis rotation is the most plausible model to account for almost all the characteristics of \kamo's lightcurves.

\section{Observations and Data Reduction}


We obtained high-cadence optical lightcurves of \kamo on UT 2026 April 20, 22, and May 9 using the \qty{8.1}{\m} Gemini North Telescope on Mauna Kea, Hawai'i, equipped with the Gemini Multi-Object Spectrograph \citep[GMOS-N;][]{2004PASP..116..425H}. The GN observations were conducted under non-sidereal tracking with a standard $r$-band filter and exposure times of \qtyrange{40}{45}{\second}. To supplement this dataset, observations were carried out on UT 2026 May 7 with the  \qty{4.3}{\m} Lowell Discovery Telescope in Happy Jack, Arizona, utilizing the Large Monolithic Imager \citep[LMI;][]{2014SPIE.9147E..2NB}. A broadband \(V\!R\) filter was used to maximize the signal-to-noise ratio of the object with \qty{40}{\second} exposure time to avoid trailing. Short exposure times were selected to provide a dense sampling cadence of \qtyrange{0.89}{1.25}{\min}, ensuring that the lightcurve variations of this fast rotator were not smoothed out during individual integration.  Despite the extreme faintness of the asteroid at an apparent magnitude of $\sim 23\text{ mag}$ during the campaign, these large-aperture facilities generally yielded excellent SNRs of \numrange{10}{15}. The detailed observational log and geometric circumstances are reported in Table~\ref{tab:obs}.

\begin{deluxetable*}{llcccccccc}[htb!]
\tablecaption{Summary of our ground-based lightcurve data of \kamo. \label{tab:obs}}
\tablehead{
\colhead{UTC\tnm{a}} &
\colhead{Facility} &
\colhead{Duration} &
\colhead{$N_{\rm obs}$\tnm{b}} &
\colhead{Typical Cadence} &
\colhead{RA\tnm{c}} &
\colhead{Dec\tnm{d}} &
\colhead{$r_h$\tnm{e}} &
\colhead{$\Delta$\tnm{f}} &
\colhead{$\alpha$\tnm{g}} \\
\colhead{} &
\colhead{} &
\colhead{} &
\colhead{} &
\colhead{(\si{\min})} &
\colhead{(\si{\deg})} &
\colhead{(\si{\deg})} &
\colhead{(\si{\au})} &
\colhead{(\si{\au})} &
\colhead{(\si{\deg})}
}
\startdata
2026-04-20T08:12:50 & GN  & \qty{43}{\min}~\qty{14}{\second} & 35 & 1.25 & 164.4840 & 40.9831  & 1.1004 & 0.198 & 56.4 \\
2026-04-22T08:34:01 & GN  & \qty{32}{\min}~\qty{16}{\second} & 28 & 1.17 & 164.0501 & 39.5015  & 1.0996 & 0.200 & 57.2 \\
2026-05-07T03:51:38 & LDT & \qty{36}{\min}~\qty{14}{\second} & 34 & 0.89 & 163.4430 & 28.4409  & 1.0909 & 0.219 & 62.6 \\
2026-05-09T07:43:51 & GN  & \qty{69}{\min}~\qty{28}{\second} & 56 & 1.24 & 163.6337 & 26.8411  & 1.0892 & 0.222 & 63.4 \\
\enddata
\tntext{a}{UTC of the start of observations.}
\tntext{b}{Number of photometric data points.}
\tntext{c}{Right Ascension (equatorial J2000).}
\tntext{d}{Declination (equatorial J2000).}
\tntext{e}{Heliocentric distance at the start of observation.}
\tntext{f}{Geocentric distance at the start of observation.}
\tntext{g}{Solar phase angle at the start of observation.}
\tablecomments{All geometric parameters correspond to the start of observations for each epoch.}
\end{deluxetable*}

Before photometric measurements, we applied standard bias and flat-field corrections to all raw images.  \kamo was uniquely identified in the field from other moving objects referencing the JPL Horizons ephemerides \citep{Horizons}.

\begin{figure*}[htb!]
\centering
\includegraphics[
    width=\textwidth,
    height=0.85\textheight,
    keepaspectratio
]{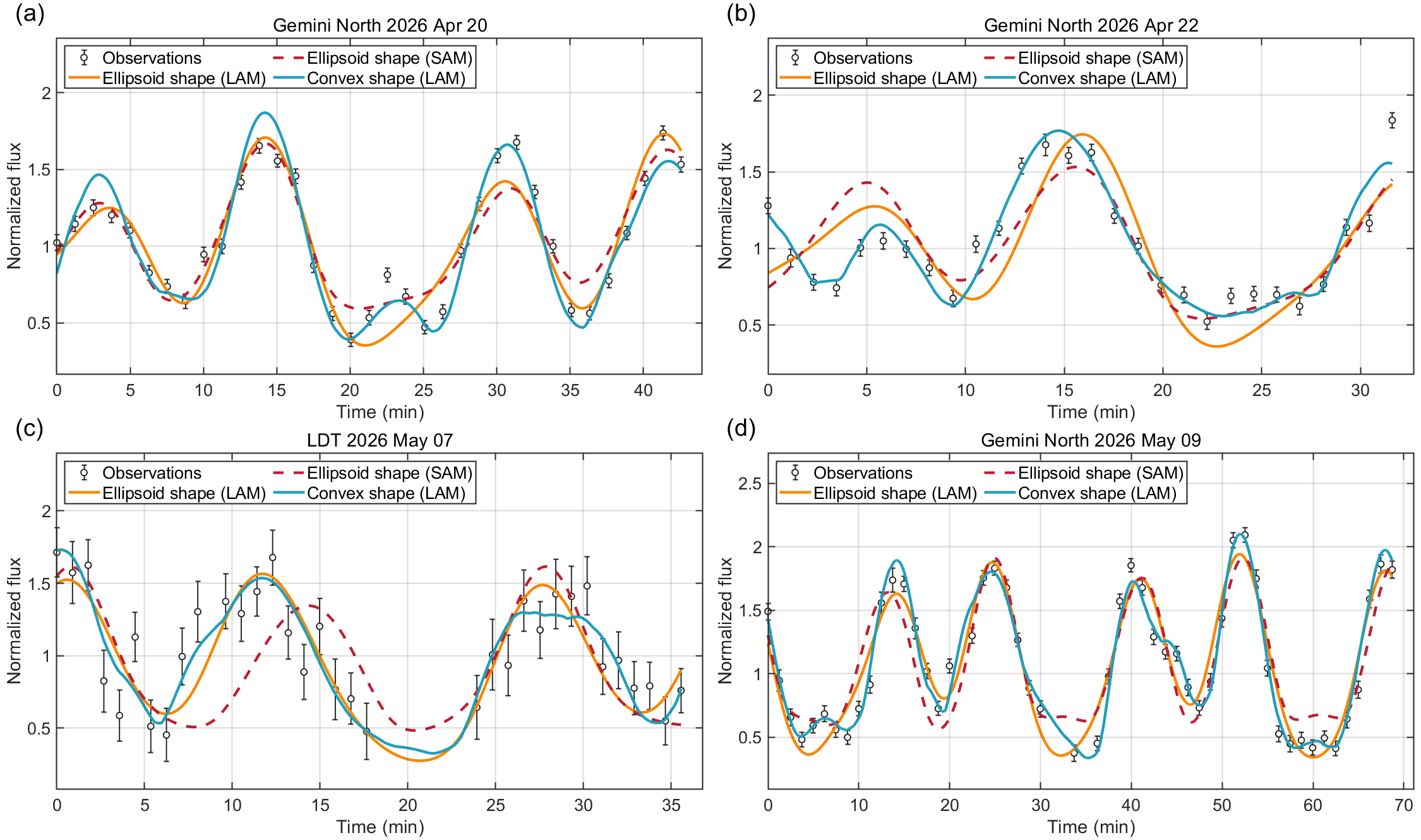}
\caption{Observed photometric lightcurves (black symbols with error bars representing the $1\sigma$ photometric uncertainties) of \kamo from GN on 2026 April 20 (a), April 22 (b), and May 9 (d), and from LDT on May 7 (c). Overplotted are the best-fit models assuming non-principal-axis rotation using a triaxial ellipsoidal shape in LAM (orange solid curve) and SAM (red dashed curve) (see \S{\ref{sec:rotation}}), and a convex shape in LAM (blue solid curve) (see \S{\ref{sec:shape}}). All lightcurves are plotted in linear flux scale and normalized to unity at their respective averages.}
\label{fig:lightcurve}
\end{figure*}

Differential aperture photometry was performed to extract the lightcurve of \kamo.  For all frames in each epoch, we chose a set of 4 to over 10 reference field stars, depending on the length of the tracked arc on the sky, from the SDSS DR16 catalog \citep{2020ApJS..249....3A} as photometric standards.  With non-sidereal tracking, the asteroid's point-like appearance allowed measurements with standard circular apertures.  However, the apertures for the field stars were dynamically adjusted in width frame by frame to accommodate the local seeing fluctuations and individual source brightness to maximize the SNR.  After subtracting the sky background, the differential magnitudes of \kamo were measured to construct the lightcurves (Fig.~\ref{fig:lightcurve}).  We verified the photometric stability of all stars across all frames in each epoch to ensure the photometric accuracy of the lightcurves.

\section{Period Analysis}
\label{sec:period_analysis}

\begin{figure*}[htb!]
\centering
\begin{subfigure}{0.49\textwidth}
    \centering
    \includegraphics[width=\linewidth]{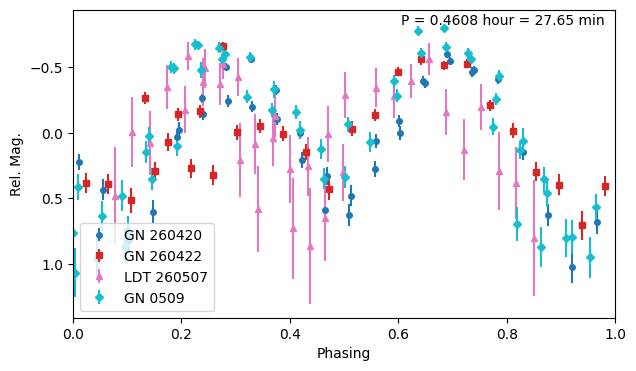}
    \caption{}
    \label{fig:phasefold_a}
\end{subfigure}
\begin{subfigure}{0.49\textwidth}
    \centering
    \includegraphics[width=\linewidth]{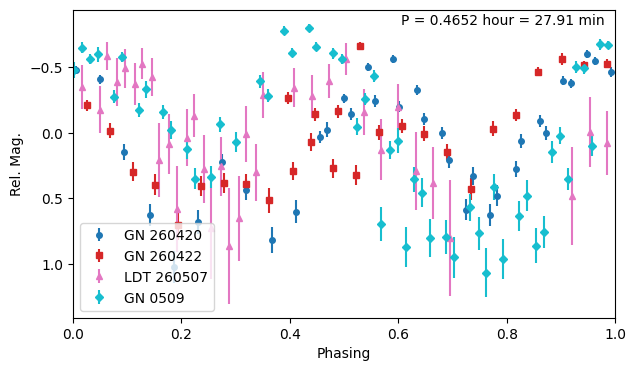}
    \caption{}
    \label{fig:phasefold_b}
\end{subfigure}
\caption{The phase folded lightcurve of \kamo with $P=\qty{27.65}{\min}$ (a) and $P=\qty{27.91}{\min}$ (b).  All lightcurves are normalized to relative magnitudes with respect to their respective medians. Error bars represent $1\sigma$ photometric uncertainties.
\label{fig:phasefold}}
\end{figure*}

First, we searched for periodicity in our newly obtained lightcurves of \kamo.  Every lightcurve was normalized to its own average, and a light time correction was also applied.  The period search was performed using the Lomb-Scargle approach \citep[e.g.,][]{2018ApJS..236...16V}.  The periodogram shows several narrow peaks with similar power levels due to aliasing from the ambiguity in the number of rotations between observing epochs, given the asteroid's fast rotation.  We therefore tested the first 10 peak periods by phasing all lightcurves together, assuming a double-peak lightcurve following previous work.  The two periods with similar peak power in the periodogram, $P=\qty{27.65(0.01)}{\min}$ and $P=\qty{27.91(0.01)}{\min}$, appear to phase all lightcurves together the best and almost equally well (Fig.~\ref{fig:phasefold}).  The error bar was estimated from the width of the corresponding peaks.  Note that $P=\qty{27.91(0.01)}{\min}$ is consistent with the most recent result \citep{2026arXiv260624017S}.  We could not distinguish between the two periods to determine which is better based on the four epochs of lightcurve used in this work.

However, we noticed that even with the best-fit phasing, the lightcurve morphologies did not fully repeat, despite the small changes of $\lesssim\qty{7}{\degree}$ in phase angle and $\lesssim$\qty{15}{\degree} in the Earth-\kamo direction over all four epochs.  Even for the May 9 lightcurve, which covers about 2.4 cycles, the mismatch is greater than the photometric uncertainties.  In particular, the extrema of adjacent cycles in this lightcurve have varying amplitudes, even though their timings align well with the best-fit period.  We therefore searched for the possible existence of periodicities other than the values that we found and reported in the literature \citep{2021ComEE...2..231S, 2026A&A...710L..35B, 2026arXiv260624017S, 2026arXiv260625512W}, but did not find any statistically significant ones.  The periodogram shows two areas of possible periodicity around $\sim\qty{17}{\min}$ and $\sim\qty{50}{\min}$, but with only about 1/3 of the peak power.  We acknowledge that the coverage of our data in the frequency domain is incomplete, especially for periods $\gtrsim\qty{2}{\hour}$.  Therefore, we could have missed the periodicity of a few days.  But whatever caused the variations in the lightcurves likely has stable periodicities.

\section{Rotational State Analysis}
\label{sec:rotation}




\subsection{Non-principal-axis rotation model}

Stable periodicity with varying amplitudes in lightcurves has been previously reported in many non-principal-axis (NPA) rotations of asteroids and comets \citep[e.g.,][etc.]{2014Icar..233...48P, 2018ApJ...856L..21B, 2013Icar..222..595B, 2020A&A...635A.137L, 2017AJ....154..136L}.  Non-principal-axis rotation can result in periodicities in asteroids' lightcurves due to the periodic repeatability of the apparent cross-section \citep{2015Icar..248..347S, 2001A&A...376..302K}.  Therefore, to explain the characteristics of \kamo's lightcurves, we adopted a force-free NPA rotation model to fit its lightcurves.

To describe the NPA rotation of \kamo, we adopted the L-convention \citep{1991Icar...93..183B, 1991Icar...93..194S} using the terminology of \citet{2015Icar..248..347S}.  The NPA rotational model of the asteroid is characterized by an angular momentum $\mathbf{L}$ that is fixed in the inertia frame, a mean precession period, $P_\phi$, of the long principal axis about $\mathbf{L}$, and the rotation (for long-axis mode, LAM) or the libration (for short-axis mode, SAM) period, $P_\psi$, of the body about its long principal axis.  Under this convention, the nutation angle $\theta$ is defined as the angle between $\mathbf{L}$ and the positive long principal axis.  For the body-fixed frame, we chose the short principal axis as the z-axis, and the long principal axis as the x-axis, and the y-axis completes the coordinate system following the right-hand rule.

In the fitting process, we first assumed a triaxial ellipsoidal shape.  We fitted all lightcurves simultaneously to optimize the 8 nonlinear model parameters: the global initial attitude quaternion $\boldsymbol{q}_0$ (3 free parameters), the direction and magnitude of the initial angular velocity $\boldsymbol{\omega}_0$ (3 free parameters), and two ellipsoid axis ratios $b/a$ and $c/a$ ($a$, $b$, and $c$ being the long-, intermediate-, and short-axis of the ellipsoid, respectively) that determine the ratios of the principal moments of inertia $I_a/I_c$ and $I_b/I_c$.  The angular momentum was not fitted independently but derived from the best-fit inertia tensor and the initial angular velocity as $\mathbf{L}_{\rm body}=\mathbf{I}\boldsymbol{\omega}_0$.  

To calculate the lightcurve of the rotating ellipsoid, we tested both the Lambertian and the Lommel-Seeliger scattering models, representing bright and dark surfaces, respectively.  The results were consistent with one another within the photometric uncertainties, demonstrating that shape and orientation dominate the lightcurves.  Given the possible high albedo of this object \citep{2026arXiv260624017S}, we adopted the Lambertian scattering model in our model fitting.

To find the best-fit rotational model, we generated sets of initial parameters that covered about \num{4e6} dynamically independent rotational states, almost uniformly sampling the entire parameter space for NPA rotation: the full $4\pi$ solid angle for the angular momentum direction, \numrange{0.2}{0.9} for axial ratios, and \qtyrange{10}{70}{\min} for the precession period, $P_\phi$, and the period of rotation, $P_\psi$.  Once the object's initial attitude and rotational parameters are set, we propagated its attitude using the force-free rigid-body rotation theory with numerical integration.  The optimization was performed using both chi-squared and particle swarm optimization approaches based on the $\chi^2$, which is defined as,
\[
\chi^2 = \sum_i{\left(\frac{F_{obs,i} - F_{mod,i}}{\sigma_i}\right)^2}
\]
We also define the weighted root-mean-squared (RMS), which represents the weighted average residual to quantify the goodness of fit,
\[
\text{RMS}=\sqrt{\frac{\chi^2}{\sum_i{\left(1/\sigma_i\right)^2}}}
\]
where $F_{obs,i}$ and $F_{mod,i}$ are the observed and modeled fluxes, respectively, both normalized; $\sigma_i$ is the photometric uncertainty; and the sum is over all data points. The resultant parameters converged to a few small regions in the parameter space.  After visually inspecting the fit, two solutions for the periods appear to fit the data equally well, corresponding to LAM and SAM, respectively, with the latter having about 10\% worse RMS than the former.  Also, for each rotational mode, two equivalent solutions exist for the direction of angular momentum, which are approximately mirrored with respect to the Sun-asteroid-Earth plane and imply opposite directions of rotation for the asteroid \citep{2006InvPr..22..749K}. Note that the mirror equivalence will be broken if the Sun, asteroid, and Earth do not move in a plane over the course of observations, which is not the case here. We then repeated the model fitting process with sets of initial parameters near the possible solutions to refine the results.  The final best-fit lightcurve model is shown in Fig.~\ref{fig:lightcurve}. The model parameters are listed in Table~\ref{tab:axis}, where we also include the mean nutation angle, $\langle\theta\rangle$, and its half amplitude, $A_{\theta}=(\theta_{\max}-\theta_{\min})/2$, which are both calculated over one complete nutation cycle.  In addition, we provided animated figures in Appendix \ref{app:ellipsoid} (Fig.~\ref{fig:animation} and Fig.~\ref{fig:animation_sam}) to illustrate the rotation of the ellipsoidal shape model together with each corresponding lightcurve for comparison.



\begin{deluxetable*}{lccccccccccc}[htb!]
\tablecaption{The best-fit rotational models of \kamo.\label{tab:axis}}
\tablehead{
\colhead{Model} &
\colhead{$\lambda$} &
\colhead{$\beta$} &
\colhead{$P_{\phi}$ (\si{\min})} &
\colhead{$P_{\psi}$ (\si{\min})} &
\colhead{$\langle\theta\rangle$} &
\colhead{$A_{\theta}$ } &
\colhead{$b/a$} &
\colhead{$c/a$} &
\colhead{$I_a/I_c$} &
\colhead{$I_b/I_c$} &
\colhead{RMS}
}
\startdata
\multicolumn{12}{c}{Ellipsoidal Shape} \\
\hline
LAM &
\qty{226}{\degree} & \qty{-39}{\degree} &
\multirow{2}{*}{27.65} & \multirow{2}{*}{50.49} &
\multirow{2}{*}{\qty{67}{\degree}} &
\multirow{2}{*}{\qty{3}{\degree}} &
\multirow{2}{*}{0.57} & \multirow{2}{*}{0.44} &
\multirow{2}{*}{0.39} & \multirow{2}{*}{0.90} & 0.178 \\
LAM Mirror &
\qty{75}{\degree} & \qty{-17}{\degree} &
& & & & & & & & 0.187 \\
SAM &
\qty{151}{\degree} & \qty{16}{\degree} &
\multirow{2}{*}{18.86} & \multirow{2}{*}{51.78} &
\multirow{2}{*}{\qty{90}{\degree}} &
\multirow{2}{*}{\qty{18}{\degree}} &
\multirow{2}{*}{0.60} & \multirow{2}{*}{0.36} &
\multirow{2}{*}{0.36} & \multirow{2}{*}{0.83} & 0.195 \\
SAM Mirror &
\qty{334}{\degree} & \qty{-33}{\degree} &
& & & & & & & & 0.206 \\
\hline
\multicolumn{12}{c}{Convex Shape} \\
\hline
LAM &
\qty{225}{\degree} & \qty{-43}{\degree} &
\multirow{2}{*}{27.90} & \multirow{2}{*}{49.66} &
\multirow{2}{*}{\qty{66}{\degree}} &
\multirow{2}{*}{\qty{3}{\degree}} &
& &
\multirow{2}{*}{0.385} & \multirow{2}{*}{0.865}& 0.106\\
LAM Mirror &
\qty{80}{\degree} & \qty{-30}{\degree} &
& & & & & & & & 0.126\\
\enddata
\tablecomments{$\langle\theta\rangle$ is \qty{90}{\degree} for SAM by definition.}
\end{deluxetable*}

\subsection{Uniqueness of model solutions}

After finding solutions that agree with the data, we assessed the uniqueness of the solutions by scanning the parameter space.  In particular, we scanned the whole sky for the direction of angular momentum, and scanned the periods near the best-fit solutions.

\begin{figure*}[htb!]
\centering
\includegraphics[width=1\textwidth]{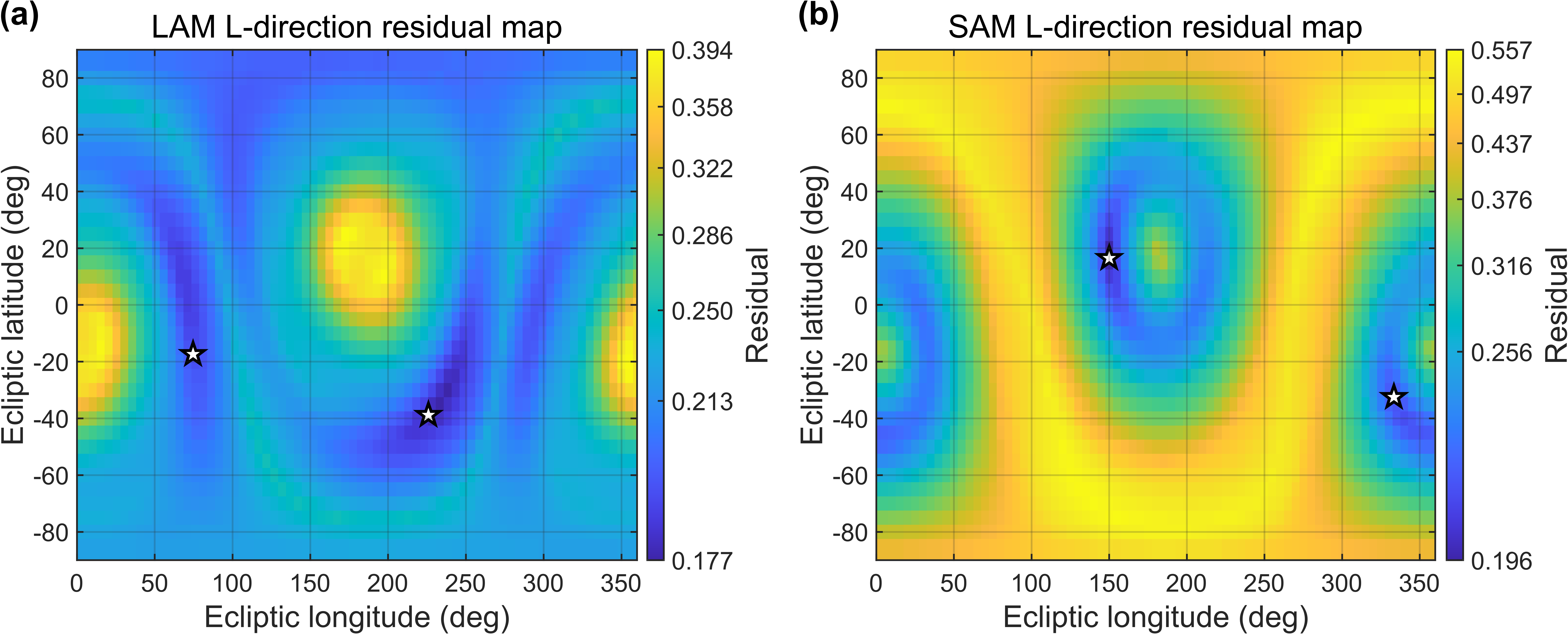}
\caption{RMS residual maps showing the solutions for the direction of angular momentum for LAM (a) and SAM (b).  The color bars indicate the RMS residuals of model fitting.  The stars mark the pair of mirrored angular momentum directions for each mode due to the symmetry about the Earth-asteroid-Sun plane.}
\label{fig:L_distribution}
\end{figure*}

For the direction of angular momentum, we fixed the direction of angular momentum in an arbitrary direction, and fitted the other parameters to find the corresponding minimum RMS.  After scanning the entire $4\pi$ solid angle for the angular momentum, it is clear that the four solutions, including LAM model and SAM model and their respective mirrored angular momentum directions, correspond to similar local minimum in RMS (Fig.~\ref{fig:L_distribution}).

For the period scan, we did it for the LAM solution as a check.  On a grid of $P_\phi$ from 26 to \qty{29}{\min} and $P_\psi$ from 48 to \qty{53}{\min}, with a step of \qty{0.05}{\min}, we initiated the ellipsoidal model with geometric scattering described by \cite{Dur.ea:25} and converged to the local minimum. The result is shown in Fig.~\ref{fig:period_scan}. The best solution, defined by having the lowest RMS, is marked with a red point, but many other local minima provide only slightly worse fits. There is a clear pattern in the separation between local minima that is related to the observational data set. The four lightcurves were observed in two blocks separated by 17 days; each block consisted of two observations, two days apart. The 2-day separation between lightcurves corresponds to about 100 cycles of $P_\phi$ and 60 cycles of $P_\psi$. Increasing or decreasing the number of cycles by one changes the periods by $\Delta P_\phi = \qty{0.27}{\min}$ and $\Delta P_\psi = \qty{0.86}{\min}$, which is twice the separation between local minima, likely because there are two maxima and minima per period (see Fig.~\ref{fig:period_scan}). The finer structure of the residual plot in the period parameter space is likely related to the 17-day separation between the observation blocks.

\subsection{Uncertainty estimate}

\begin{figure*}[htb!]
\centering
\includegraphics[width=0.85\textwidth]{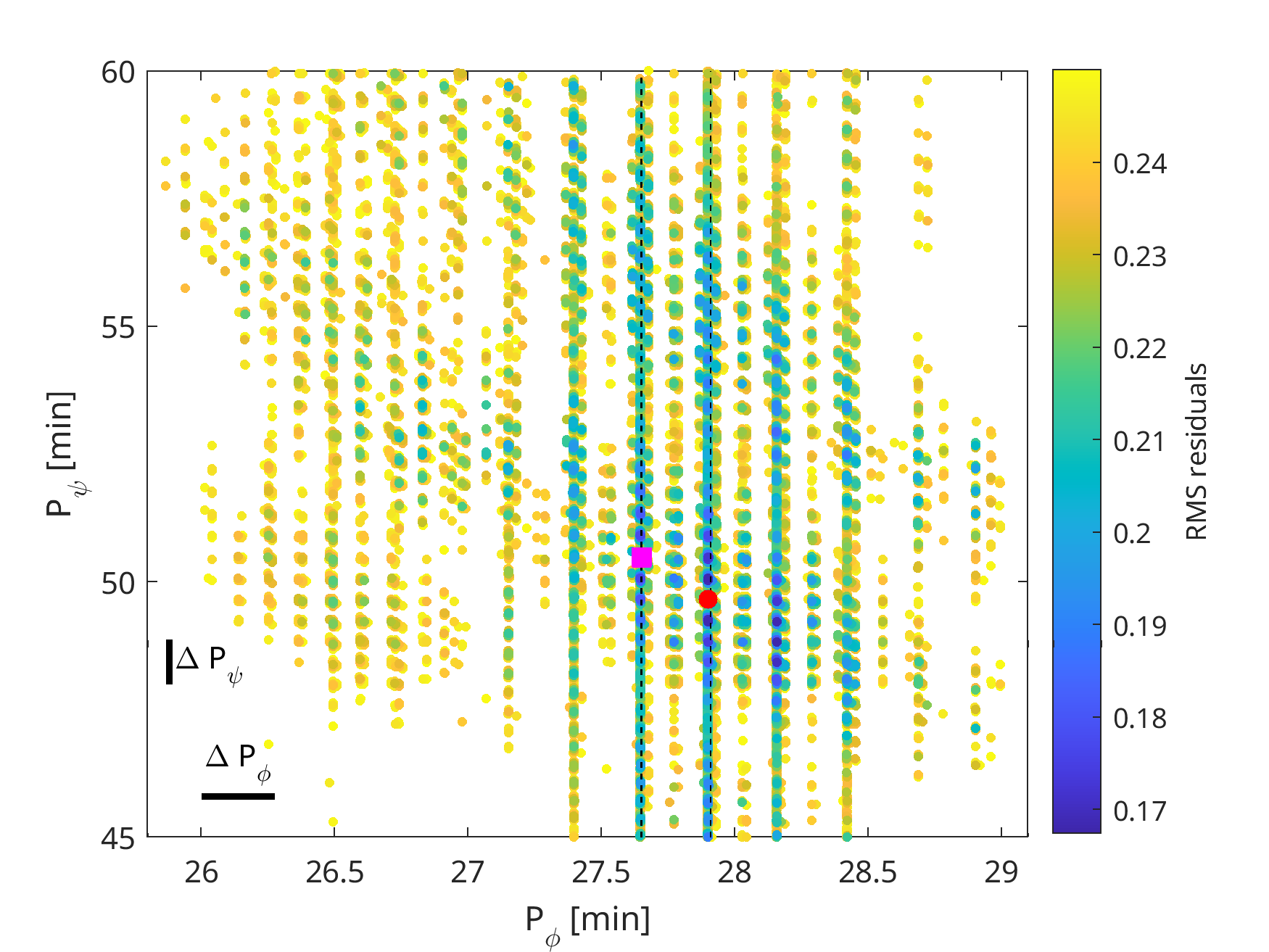}
\caption{Period scan for ellipsoidal shape.  The colorbar marks the model RMS. For clarity, points with RMS values above 0.25 are not plotted. The best-fit value with the lowest RMS is marked with a red point and corresponds to the values in Table~\ref{tab:axis}.  The pink square marks the second best solution.  The two dashed vertical lines mark the two best periods found by the period analysis described in \S\ref{sec:period_analysis}. The thick horizontal and vertical lines show the separations $\Delta P_\psi$ and $\Delta P_\phi$ between local minima (see the main text).}
\label{fig:period_scan}
\end{figure*}

We estimated the uncertainty of the periods using multiple approaches.  For a sinusoidal lightcurve with photometric uncertainties of $\sigma_m$, the phasing uncertainty $\sigma_\phi$ in one cycle is on the order of $\sigma_m/A$, where $A$ is the half amplitude of the lightcurve.  For our data, $\sigma_m\sim\qty{0.08}{mag}$, and $2A\sim\qty{1.6}{mag}$, giving $\sigma_\phi\sim\qty{0.1}{cycle}\sim\qty{2.7}{\min}$.  With about \qty{100}{cycles} between the epochs, the uncertainty of $P_\phi$ shrinks to $\sim\qty{0.03}{\min}$.  This is consistent with the width of the peaks in Fig.~\ref{fig:period_scan} for $P_\phi$.  With the same rationale, the uncertainty of $P_\psi$ is $\sim\qty{0.08}{\min}$.

Uncertainty in the direction of angular momentum can be estimated based on the RMS distribution map (Fig.~\ref{fig:L_distribution}).  Experiments suggested that once RMS increases to about $1.2\times$ the minimum RMS, then the model cannot fit the data.  Taking this value to define the region of uncertainty, we consider that the direction of angular momentum is uncertain by about $\pm\qtyrange{20}{30}{\degree}$ in both ecliptic latitude and ecliptic longitude for the LAM models.


\subsection{Model quality assessment}

Overall, the NPA rotation models provide a good fit to the observed lightcurves, especially for LAM model.  For example, the LAM solution has all extrema in the lightcurves correspond to specific orientations of the ellipsoid shape (Fig.~\ref{fig:animation}).  Specifically, the lightcurve maximum corresponds to the time when the ellipsoid is viewed approximately along the shortest axis (Z-axis), so the maximum cross-section of the ellipsoid (X-Y plane) faces the Earth.  However, due to the rotation, some maxima correspond to viewing aspects off the Z-axis by various angles, explaining the varying magnitudes of the maxima from one rotation to the next.  On the other hand, the primary minimum of the lightcurve corresponds to the time when the longest axis (X-axis) is pointing directly towards or away from the Sun, whereas the secondary minimum corresponds to the time when the X-axis is about \qtyrange{30}{40}{\degree} away from the Sun.  Thus, the varying orientations of the X-axis from one rotation to the next caused the varying magnitude of the minimum.


We also noticed that some lightcurve maxima show appreciable discrepancies with the model.  In particular, both the observed maxima at about \qty{30}{\min} in the April 20 lightcurve and at about \qty{51}{\min} in the May 9 lightcurve are higher than the model predictions.  Examining the orientations of the ellipsoid shape model, we realized that the inconsistent maxima always correspond to a viewing aspect from -Z axis, whereas the consistent maxima always correspond to a viewing aspect from the +Z axis.  This indicates that the deviation is likely caused by local topography or albedo features specifically tied to that particular viewing angle.

The SAM model models result a slightly worse fit to the data.  In terms of the model lightcurve (Fig.~\ref{fig:lightcurve}), the fit is worse than LAM in both the lightcurve extrema and the phasing.  For example, for the LDT data, the SAM model lightcurve shows a significant phase shift from the peak to minimum between \qty{5}{\min} and \qty{20}{\min}.  This is consistent with the $\sim10\%$ worse RMS than the LAM model.  The fact that we cannot find a satisfactory convex shape inversion for SAM is also an indication that LAM is a preferable solution (see \S\ref{sec:shape}).

Given that the NPA rotation models explain almost all the characteristics of our observed lightcurves, we consider them the most plausible rotational models for \kamo.  A stable spin about the shortest axis cannot reconcile the varying shapes of the lightcurves from cycle to cycle.  However, the lightcurve data cannot uniquely determine the exact rotational mode or the direction of angular momentum, although the LAM model is slightly preferable.  The high-resolution imaging from China's Tianwen-2 mission will soon provide the ultimate test of our models and determine the exact rotational parameters.

\section{Convex Shape Inversion}
\label{sec:shape}

Based on the rotation model derived above, we applied lightcurve inversion \citep{2001A&A...376..302K} to derive a convex shape model for \kamo. The kinematic description of the complex rotation started with the four sets of ellipsoidal model parameters listed in Table~\ref{tab:axis} for both mirror solutions of both LAM and SAM.  The shape model is described as a convex polyhedron. The fitting process simultaneously adjusted the shape model and the rotational parameters to fit all four lightcurves.  This is a more realistic approximation, but it still inevitably deviates from the true, presumably nonconvex, shape.

\begin{figure*}[!thb]
\centering
\includegraphics[width=0.8\textwidth, trim=2cm 6.2cm 1.5cm 0.8cm, clip]{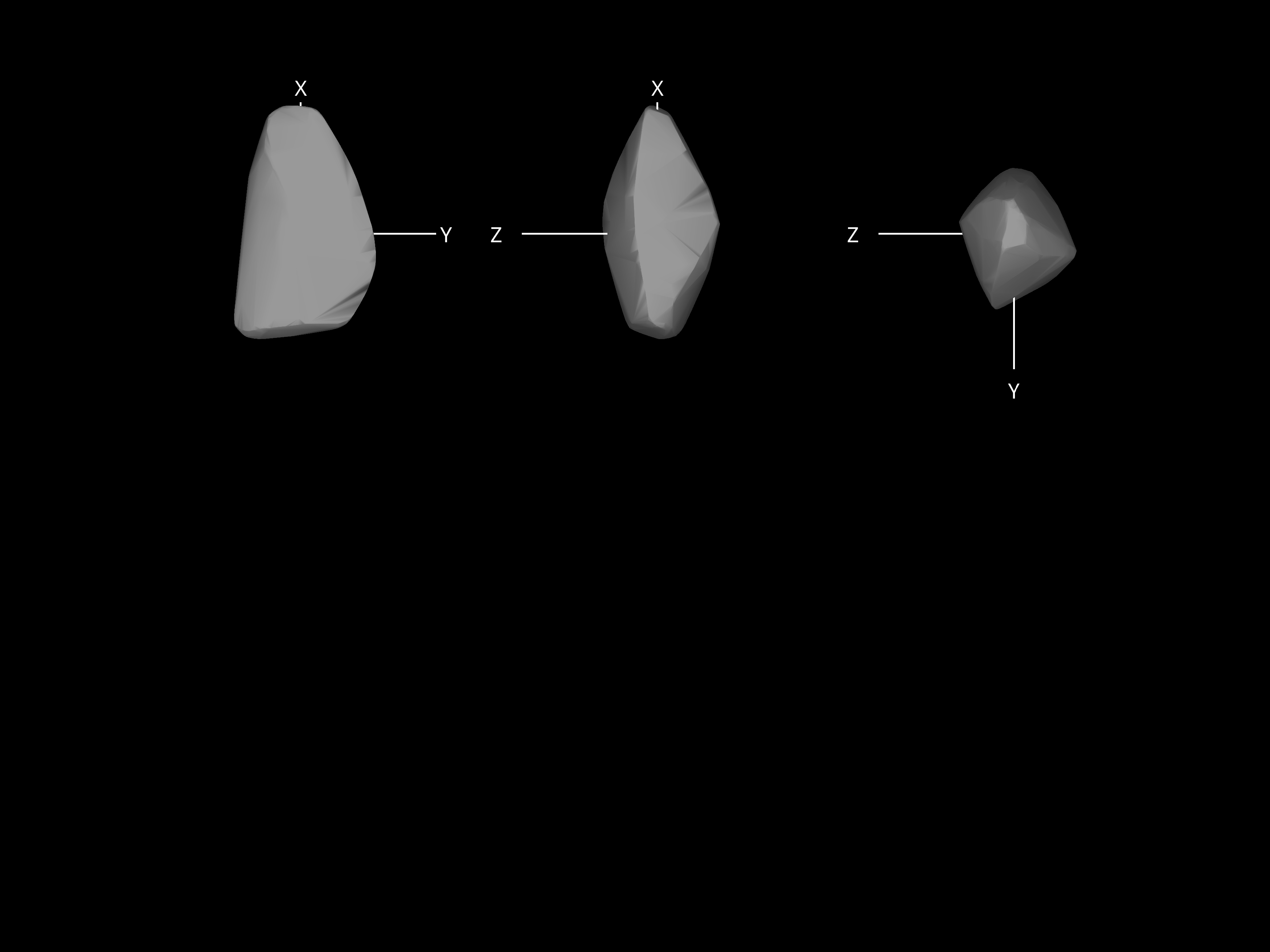}
\vspace{0cm}
\includegraphics[width=0.8\textwidth, trim=2cm 6.2cm 1.5cm 0.8cm, clip]{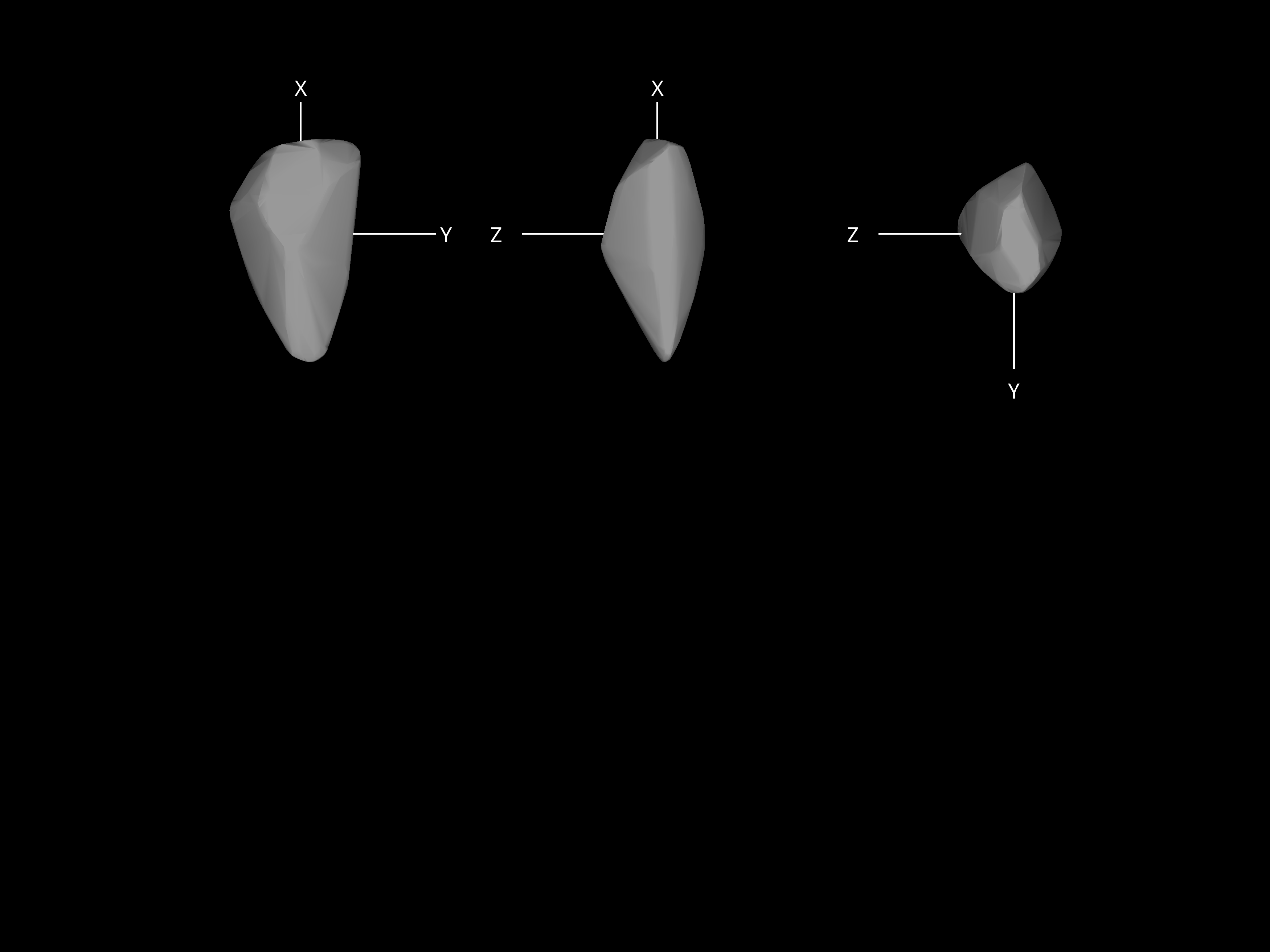}
\caption{Convex shape models obtained from lightcurves based on the rotational model described in Table~\ref{tab:axis}. The top panel show the model with the angular momentum direction ($\lambda$, $\beta$) = (\qty{225}{\degree}, \qty{-43}{\degree}), and the bottom panel show the mirror solution with ($\lambda$, $\beta$) = (\qty{80}{\degree}, \qty{-30}{\degree}).  The convex shape models are downloadable from the online journal in a triangulated plate format.}
\label{fig:shape_model}
\end{figure*}

With this process, we were able to derive the convex shape models for both mirror solutions of LAM, with the best-fit lightcurves shown in Fig.~\ref{fig:lightcurve}, and the corresponding shape models in Fig.~\ref{fig:shape_model}.  The resulting model lightcurves have overall characteristics similar to the ellipsoidal model but provide a better fit to the detailed shape, especially near the extrema.  The best-fit rotational parameters are close to the initial values adopted from the ellipsoidal model and are listed in Table~\ref{tab:axis}.  Given that our lightcurve data can only constrain the moments of inertia $I_a/I_c$ and $I_b/I_c$ to several tens of percent by the model, the values of 0.375 and 0.898 calculated from the convex shape model assuming a uniform density are consistent with those listed in Table~\ref{tab:axis}.  Therefore, this convex shape model is self-consistent with the rotational state, even if the kinematic parameters $I_a$ and $I_b$ are not coupled with the shape model during inversion.  On the other hand, for the SAM rotational parameters, we could not find any convex shape inversion that satisfactorily fit all four lightcurves simultaneously.

Note that in the lightcurve inversion, we used Hapke's scattering model with parameters fixed at values given by \cite{Li.ea:15} for E-type asteroid \v{S}teins \citep{Spj.ea:12}. This choice was motivated by the recent result of \citet{2026arXiv260624017S}, who analyzed JWST observations and concluded that the surface of \kamo is plausibly similar to E-type asteroids.  However, the values of Hapke's parameters, or even the choice of light-scattering model, play only a marginal role in the inversion, as they mainly affect the shape model, not the spin parameters.

\vspace{1.5em}

\section{Discussions}

\subsection{Non-principal-axis rotation vs. simple rotation}

The observed lightcurves exhibit a stable photometric period of approximately \qty{28}{\min}, consistent with previous measurements. However, a closer inspection reveals cycle-to-cycle variations in both the peak brightness and the detailed shape, especially near the extrema.  Such behavior is difficult to reconcile with a simple principal-axis rotation with a fixed pole orientation.  Our model, which combines photometric scattering geometry with rigid-body dynamics, can reproduce both the stable recurrence timescale and the non-repeating shape for the four observed lightcurve segments simultaneously.  The resulting solution indicates that the optical data are highly consistent with the behavior expected from a freely precessing, elongated body.

We emphasize that the presence of a stable photometric period does not contradict the NPA rotation. In a torque-free precessing rotation of a rigid body, the angular momentum vector remains fixed in inertial space, and the attitude motion is governed by a few frequencies associated with spin, precession, and nutation. The observed brightness does not require the full 3D attitude to repeat exactly. Rather, it primarily reflects the recurrence of the projected and illuminated cross section of the body as viewed from Earth. In our best-fit solution, the dominant photometric period of $\sim\qty{28}{\min}$ is closest to the long-axis precession period $P_\phi = \qty{27.65}{\min}$. This is consistent with the above analysis that the photometric recurrence timescale is mainly controlled by the precession and does not require exact repetition of the full 3D attitude, especially for a highly elongated shape like \kamo's. The non-commensurability among the spin, precession, and nutation frequencies naturally implies that the lightcurve extrema recur at a stable period while their amplitudes and detailed shapes vary from cycle to cycle.  This result is also compatible with recent JWST observations of \kamo, which reported systematic variations in lightcurve amplitude under similar observing geometries \citep{2026arXiv260624017S}.

However, one important question is why none of the previous lightcurve observations reveal the NPA rotation of \kamo.  We think that the combination of limited temporal coverage and limited SNR is the main reason.  The previous lightcurve data reported by \citet{2021ComEE...2..231S} and those from the Minor Planet Center covered at most $\sim\qty{40}{\min}$ each in any epoch.  The Keck lightcurve data reported by \citet{2018DPS....5050504F} covered a total continuous duration of $\sim\qty{100}{\min}$ in $V$, $R$, and $I$ filters with $\sim\qty{30}{\min}$ each.  However, we did not see a period analysis based on those data.  Therefore, the period analysis based on each individual lightcurve could not reveal the \qty{50}{\min} period.  The relatively low SNR of all the data (except for Keck data) compared to ours, together with the changing observing geometry that caused variations in both the overall brightness and the shape of the lightcurve, limited the ability to recognize the lightcurve variations as an indicator of NPA rotation.

For this reason, the previous shape models of \kamo derived using the lightcurve inversion technique and its variations, based on the historical data obtained in 2017 and 2018, all assumed a simple principal-axis rotation \citep{2026A&A...710L..35B, 2025A&C....5100925Z, 2021Icar..35714249L, 2026arXiv260625512W}.  The limited coverage of the observing aspects of previous data further limited the reliability of the shape model derivation.  It is therefore not surprising that all previously derived shape models and the pole orientations are inconsistent with one another and with the NPA rotation models that we derived here.

\subsection{Implications of non-principal-axis rotation}

The origin of \kamo's excited rotation offers important clues about the origin and evolutionary history of this Earth quasi-satellite.  Two leading scenarios about the dynamic origin of \kamo have received broad attention: (1) produced as lunar impact ejecta \citep{2021ComEE...2..231S, 2024NatAs...8..819J, 2026Innov...701183Z}; and (2) delivered from the main asteroid belt \citep{2026ApJ...997L..49L, zhang_tianwen-2_2026}.  In terms of internal structure, neither scenario uniquely requires nor excludes either a rubble-pile or monolithic interior. The NPA rotational state can provide an additional constraint: If the present NPA state represents a primordial spin state inherited from \kamo's formation or ejection, then the object may be more consistent with a monolithic or mechanically coherent fragment rather than with a highly dissipative rubble pile.  This is because internal dissipation associated with NPA rotation is expected to damp the excited rotational state on a timescale of \qtyrange{e4}{e5}{yr} for rubble piles \citep[c.f., e.g.,][]{2002aste.book..113P, 2002aste.book..517P}, making it difficult for a primordial NPA state to survive the possible dynamic age of \kamo as an Earth co-orbiter \citep{2016MNRAS.462.3441D, 2021AJ....162..227F}.  In contrast, a monolithic or more rigid body, with lower internal damping efficiency, could preserve such a state much longer, possibly over \qty{e8}{yr}.  We also note that the timescale for all available data of \kamo is about \qty{10}{yr}, which is several orders of magnitude shorter than the shortest realistically expected damping timescale.  Therefore, it is impossible to derive a physically meaningful lower limit for the structural parameter for \kamo based on the observational data.

Other mechanisms could also initiate or contribute to the excited rotation of \kamo after its formation, such as YORP torque \citep[etc.]{2007Icar..191..636V, 2015MNRAS.449.2489B}, tidal interaction with planets during close encounters \citep{2000Icar..147..106S}, rotational fission \citep{2011Icar..214..161J}, etc.  Future high-resolution imaging and shape characterization by Tianwen-2 will be essential for determining the internal structure, composition, and surface morphology of \kamo, helping us answer the questions about the origin of this interesting asteroid and its rotational state.

\section{Conclusions}

To conclude, we obtained multi-epoch, high-cadence, high-SNR photometric lightcurves from GN and LDT.  The lightcurves are consistent with the most probable period of $P=\qty{27.65(0.01)}{\min}$ or $P=\qty{27.91(0.01)}{\min}$, although some other periods cannot be fully excluded due to significant aliasing.  However, the amplitude and detailed shape of lightcurves vary from cycle to cycle, contradicting a simple principal-axis rotation scenario.

Assuming an NPA rotation and a triaxial ellipsoidal model, we were able to fit almost all characteristics of the lightcurves. Four solutions exist, including both LAM and SAM, as well as their mirrored directions of angular momentum due to the inherited ambiguity in fitting disk-integrated brightness (Table~\ref{tab:axis}), although LAM is slightly preferable due to the slightly lower model residual.  All the solutions require elongated shapes with axial ratios of roughly 2.5:1.7:1.  In both LAM and SAM, $P_\psi$ is close to \qty{51}{\min}, but the precession periods are $P_\phi=\qty{27.65}{\min}$ and \qty{18.86}{\min}, respectively.

Based on the NPA rotational models, we performed lightcurve inversion and derived a convex shape model for LAM, but could not find a satisfactory model for SAM.  The associated rotational parameters and the moment of inertia are close to those derived with the ellipsoidal model.  The resulting model lightcurve provided a better fit to the details of the lightcurve morphology than the ellipsoidal shape.  The shape model inversion process further suggested that LAM is more likely to represent the rotational state of \kamo.

The NPA rotation of \kamo provides additional constraints on its dynamic history or internal structure.  The excited rotation of \kamo could be inherited from its ejection from the parent body if its internal dissipation is weak, or it could be induced or affected by mechanisms such as YORP, tidal interactions with planets, and rotational fission, etc.  Tianwen-2's exploration of \kamo will provide the ultimate test of our rotational models and help constrain its origin and the internal structure of \kamo.

\begin{acknowledgments}

This work is based on observations obtained at the international Gemini Observatory (Program ID: GN-2026A-FT-210; PI: Yang), a program of NSF NOIRLab, which is managed by AURA under a cooperative agreement with the U.S. National Science Foundation on behalf of the Gemini Observatory partnership. The Gemini data were obtained from the Gemini Observatory Archive. The authors recognize and acknowledge the significant cultural role and reverence that Maunakea has within the Native Hawaiian community. Some data used in this research are obtained at the Lowell Discovery Telescope (LDT). Lowell Observatory is a private, non-profit institution dedicated to astrophysical research and public appreciation of astronomy and operates the LDT in partnership with Boston University, the University of Maryland, the University of Toledo, Northern Arizona University, and Yale University. Partial support of the LDT was provided by Discovery Communications. LMI was built by Lowell Observatory using funds from the National Science Foundation (AST-1005313). This research has made use of NASA's Astrophysics Data System and the JPL Horizons On-Line Ephemeris System provided by the Solar System Dynamics Group of the Jet Propulsion Laboratory, a federally funded research and development center managed for NASA by Caltech.  J.-Y. Li acknowledges the support by the 2024 Xinjiang Autonomous Region Tianchi Talent Program and by Natural Science Foundation of Xinjiang Uygur Autonomous Region No. 2025D01E62. J.~\v{D}urech was supported by the Czech Science Foundation grant 25-16789S. The authors thank the anonymous referee for the constructive suggestions that have helped improve this manuscript.

\end{acknowledgments}

\begin{contribution}

XS and ZS performed the modeling and contributed to the writing of the paper.
HT led data curation and contributed to the writing of the paper.
BY secured the telescope time, reduced the GN data, and reviewed the paper.
JD led the lightcurve inversion, contributed to the rotation modeling, and reviewed the paper.
NM, SH, and AT performed observations and reviewed the paper.
YY led the interpretation of the results and contributed to the writing of the paper.
JYL led the conceptualization, data analysis, interpretation, and writing of the paper.


\end{contribution}

%
\facilities{Gemini:Gillett (GMOS-N), LDT (LMI)}

\software{astropy \citep{2013A&A...558A..33A,2018AJ....156..123A,2022ApJ...935..167A},
       sbpy \citep{2019JOSS....4.1426M}
          }


\appendix

\section{Ellipsoidal Rotational Model}
\label{app:ellipsoid}
\restartappendixnumbering

\begin{figure*}[htb!]
\centering
\begin{subfigure}{0.5\textwidth}
    \begin{interactive}{animation}{LAM_GN_0420.mp4}
    \centering
    \includegraphics[width=0.9\linewidth]{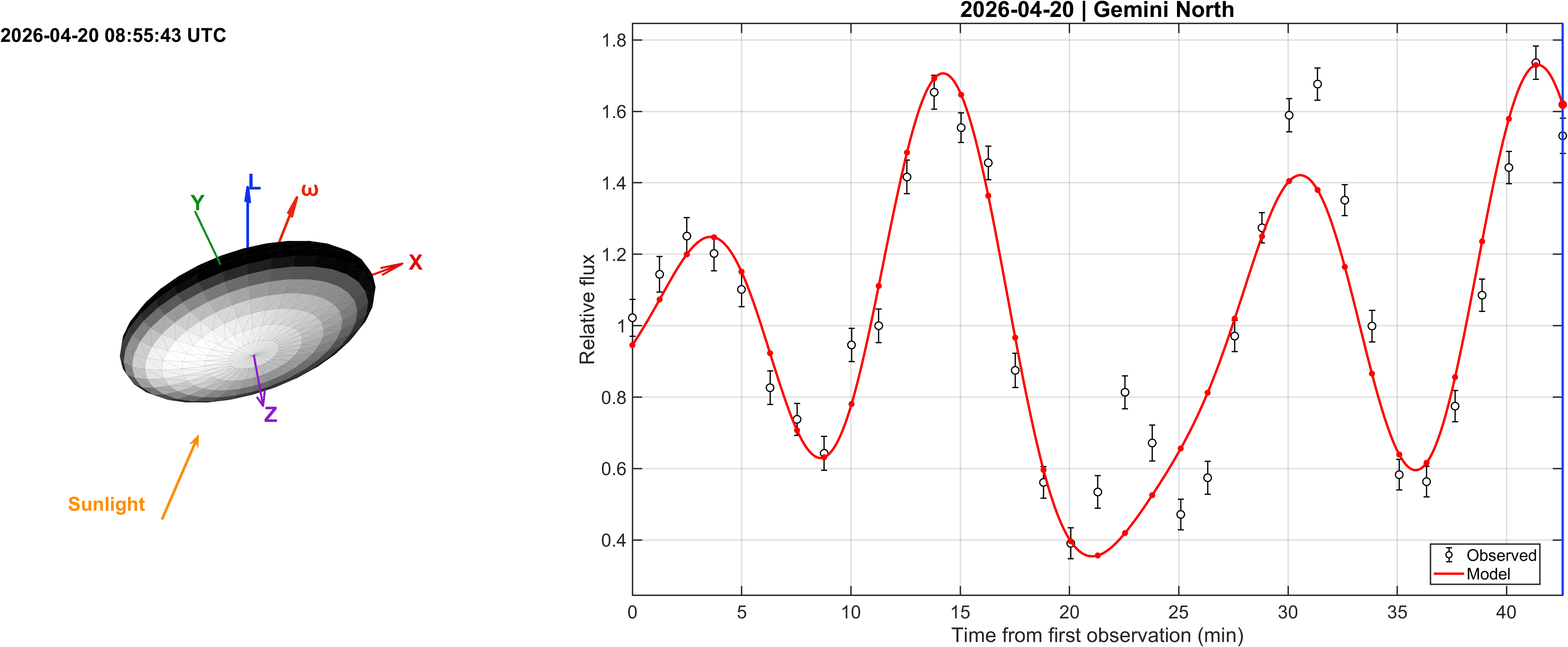}
    \caption{}
    \end{interactive}
\end{subfigure}
\hfill
\begin{subfigure}{0.5\textwidth}
    \begin{interactive}{animation}{LAM_GN_0422.mp4}
    \centering
    \includegraphics[width=0.9\linewidth]{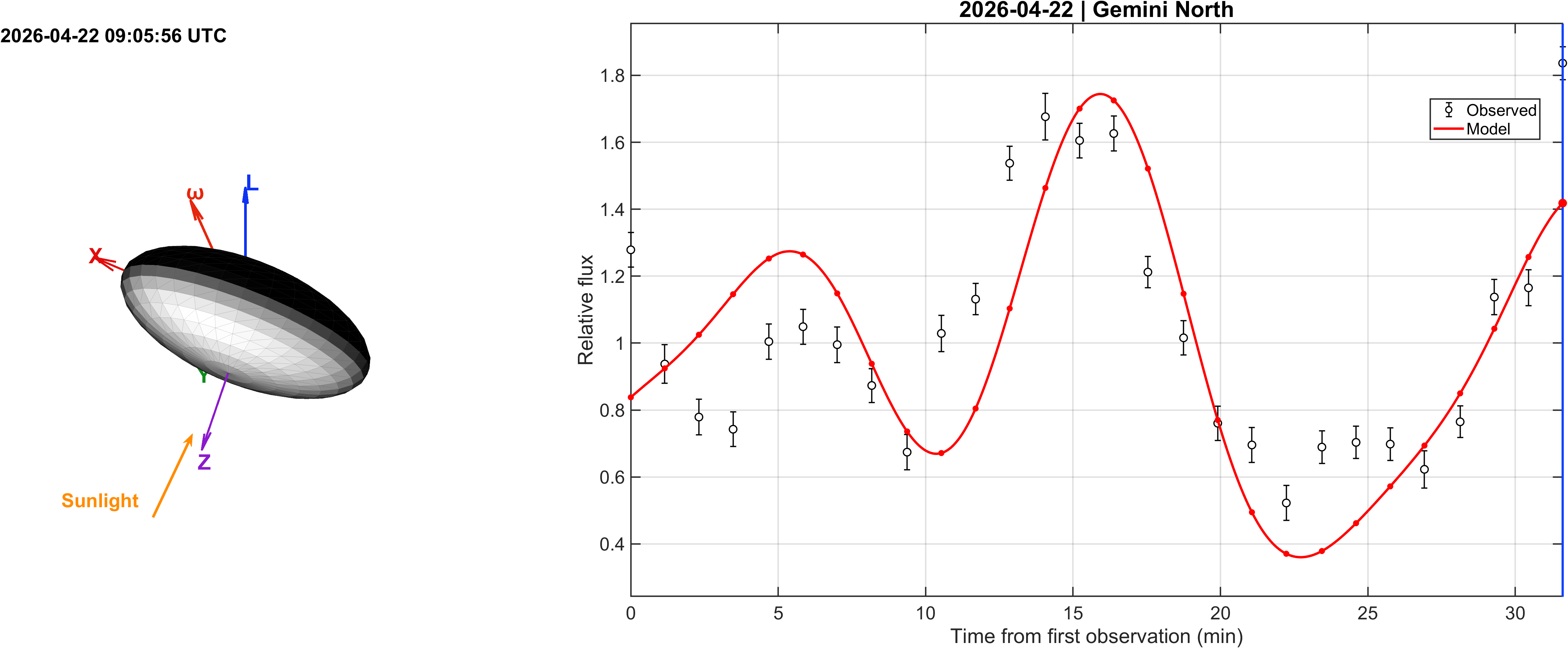}
    \caption{}
    \end{interactive}
\end{subfigure}
\vspace{0cm}
\begin{subfigure}{0.5\textwidth}
    \begin{interactive}{animation}{LAM_GN_0509.mp4}    
    \centering
    \includegraphics[width=0.9\linewidth]{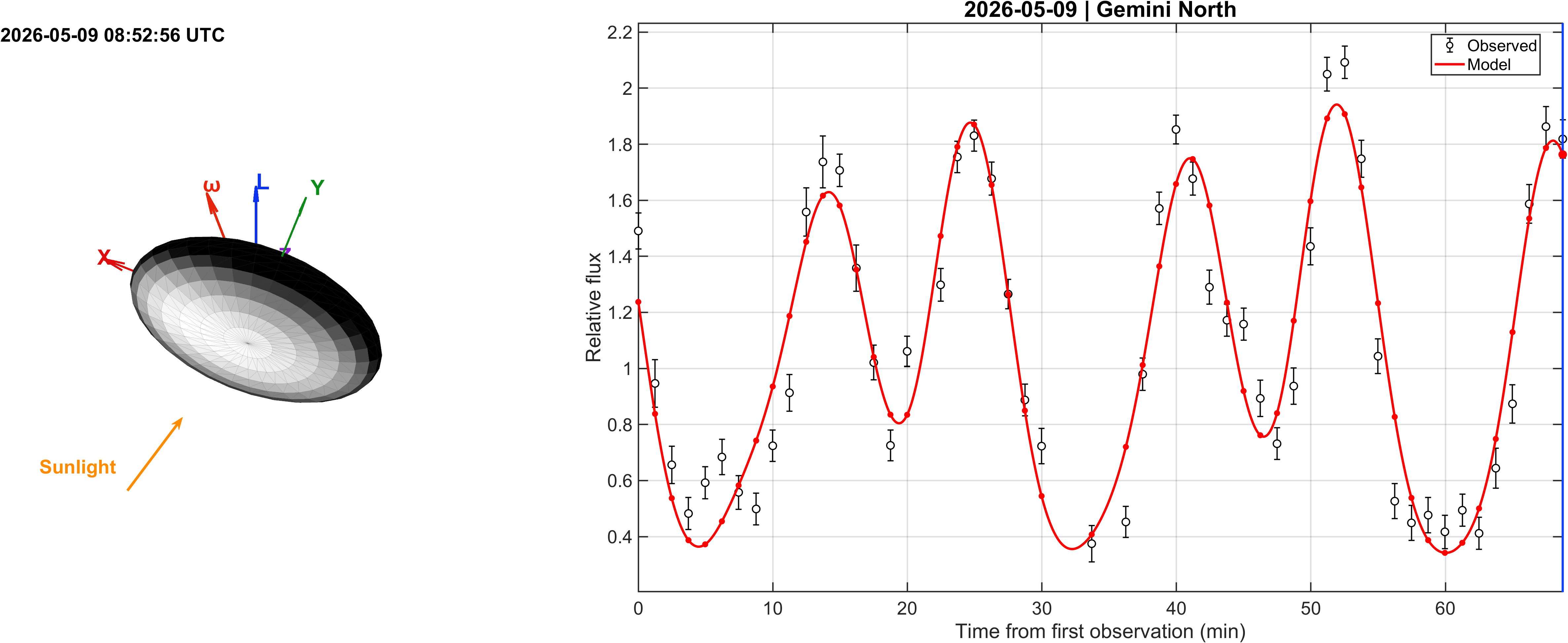}
    \caption{}
    \end{interactive}
\end{subfigure}
\hfill
\begin{subfigure}{0.5\textwidth}
    \begin{interactive}{animation}{LAM_LDT_0507.mp4}
    \centering
    \includegraphics[width=0.9\linewidth]{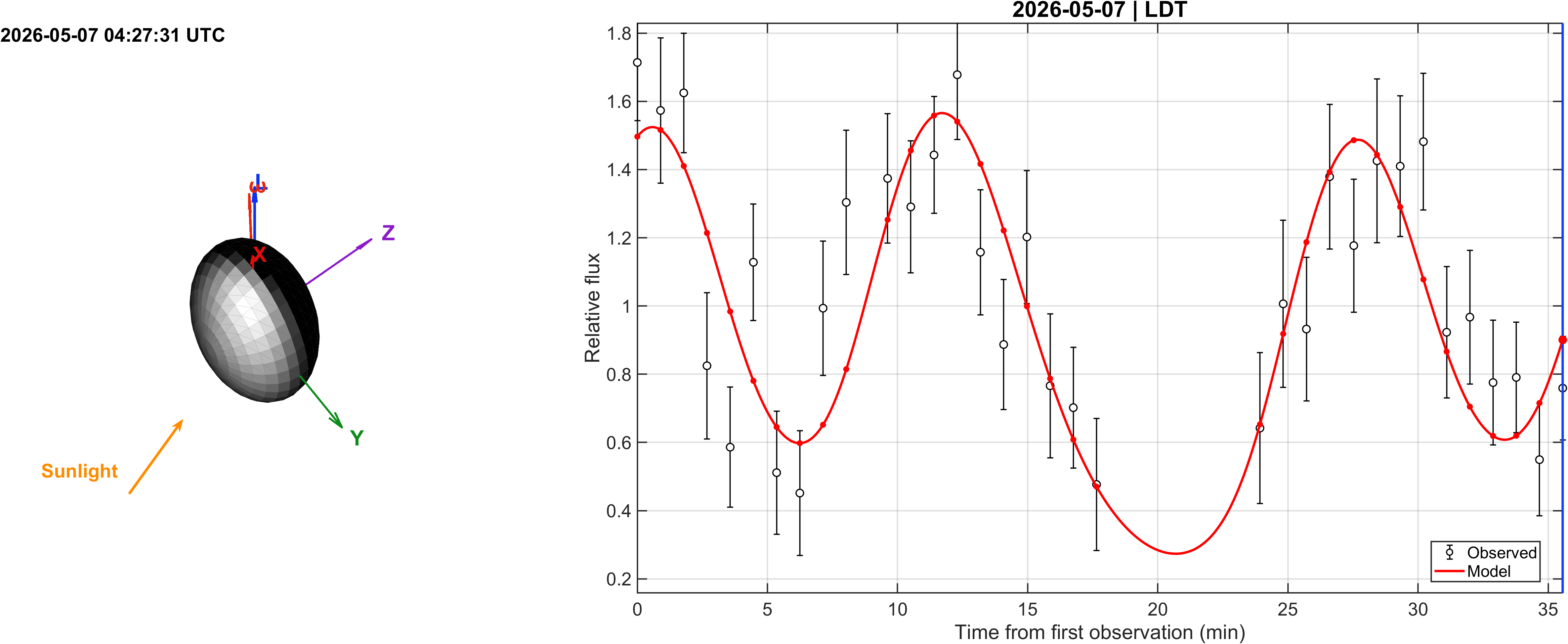}
    \caption{}
    \end{interactive}
\end{subfigure}
\caption{Animations showing the LAM rotational model and the fit to lightcurves from April 20 (a), April 22 (b), May 9 (c), and May 7 (d), respectively. The angles between the line of sight and the angular momentum vector $\mathbf{L}$ are $100^\circ$, $99^\circ$, $88^\circ$, and $89^\circ$ in the four epochs, respectively.  Therefore, the angular momentum is almost within the image plane.}
\label{fig:animation}
\end{figure*}

\begin{figure*}[htb!]
\centering
\begin{subfigure}{0.5\textwidth}
    \begin{interactive}{animation}{SAM_GN_0420.mp4}
    \centering
    \includegraphics[width=0.9\linewidth]{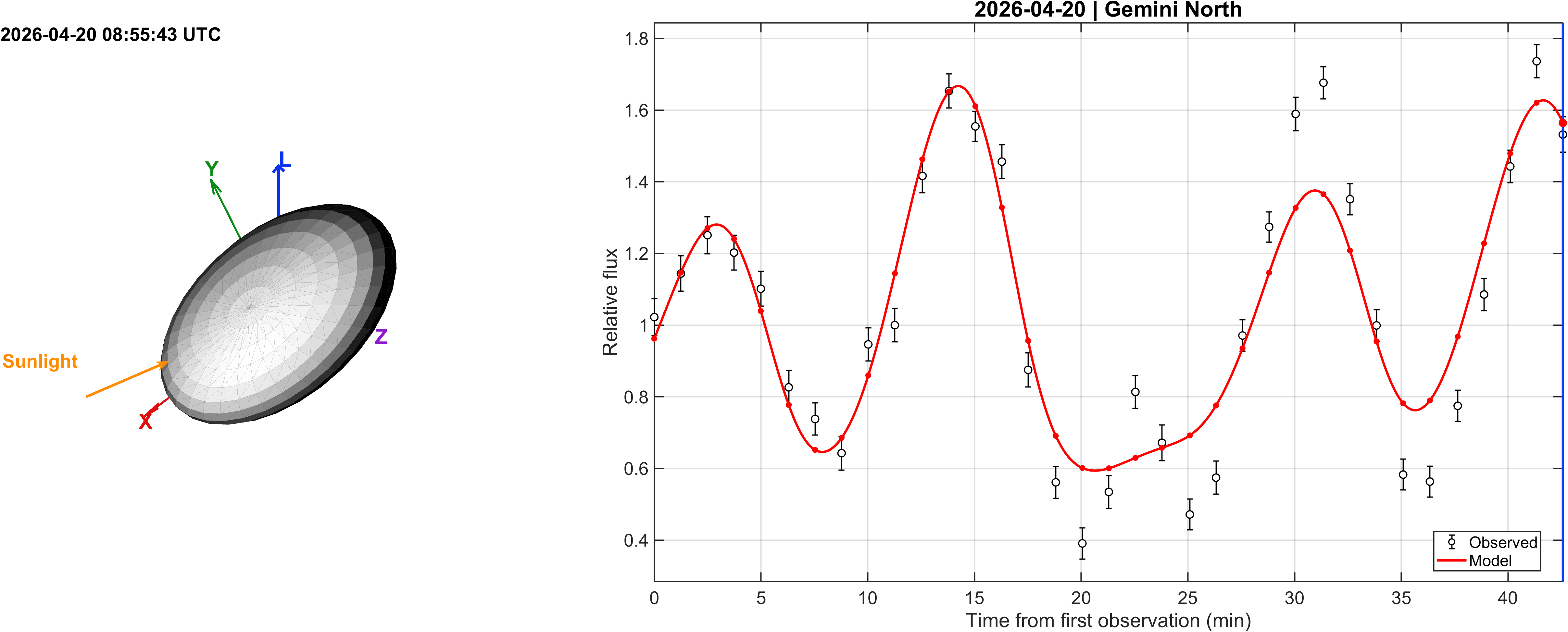}
    \caption{}
    \end{interactive}
\end{subfigure}
\hfill
\begin{subfigure}{0.5\textwidth}
    \begin{interactive}{animation}{SAM_GN_0422.mp4}
    \centering
    \includegraphics[width=0.9\linewidth]{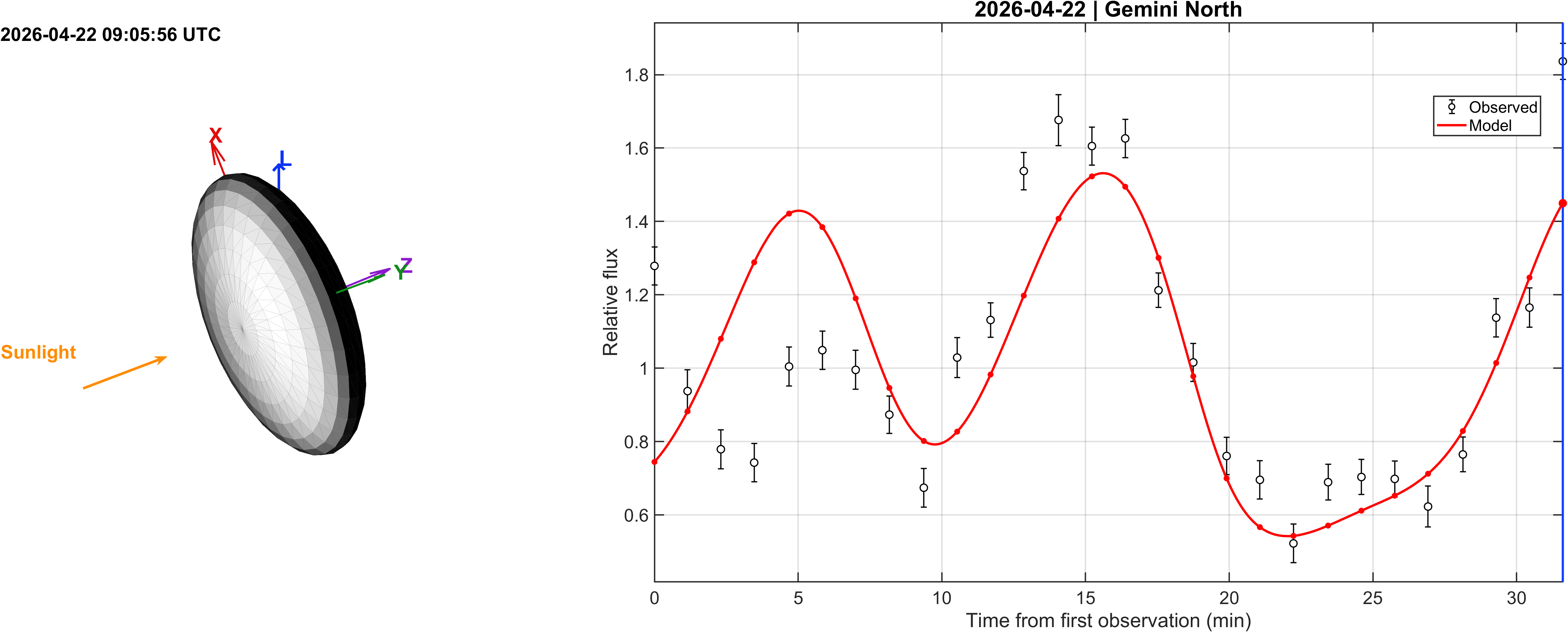}
    \caption{}
    \end{interactive}
\end{subfigure}
\vspace{0cm}
\begin{subfigure}{0.5\textwidth}
    \begin{interactive}{animation}{SAM_GN_0509.mp4}    
    \centering
    \includegraphics[width=0.9\linewidth]{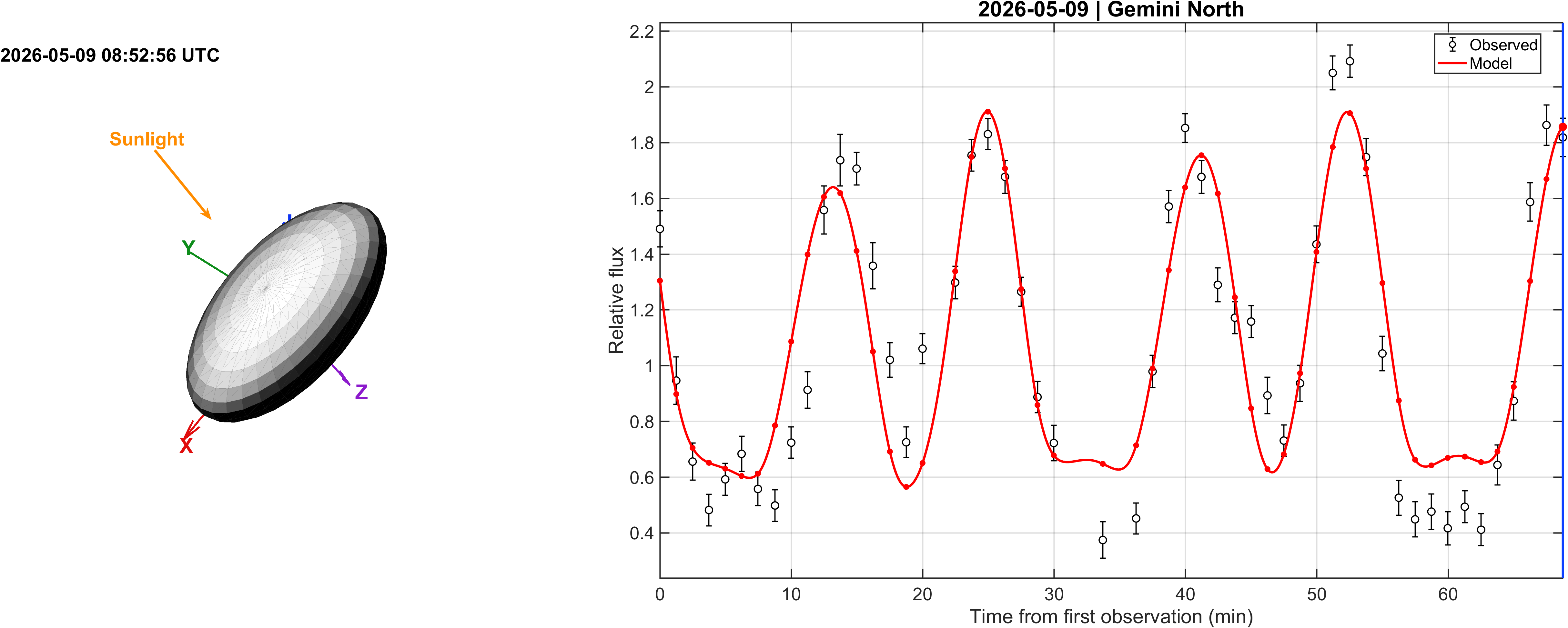}
    \caption{}
    \end{interactive}
\end{subfigure}
\hfill
\begin{subfigure}{0.5\textwidth}
    \begin{interactive}{animation}{SAM_LDT_0507.mp4}
    \centering
    \includegraphics[width=0.9\linewidth]{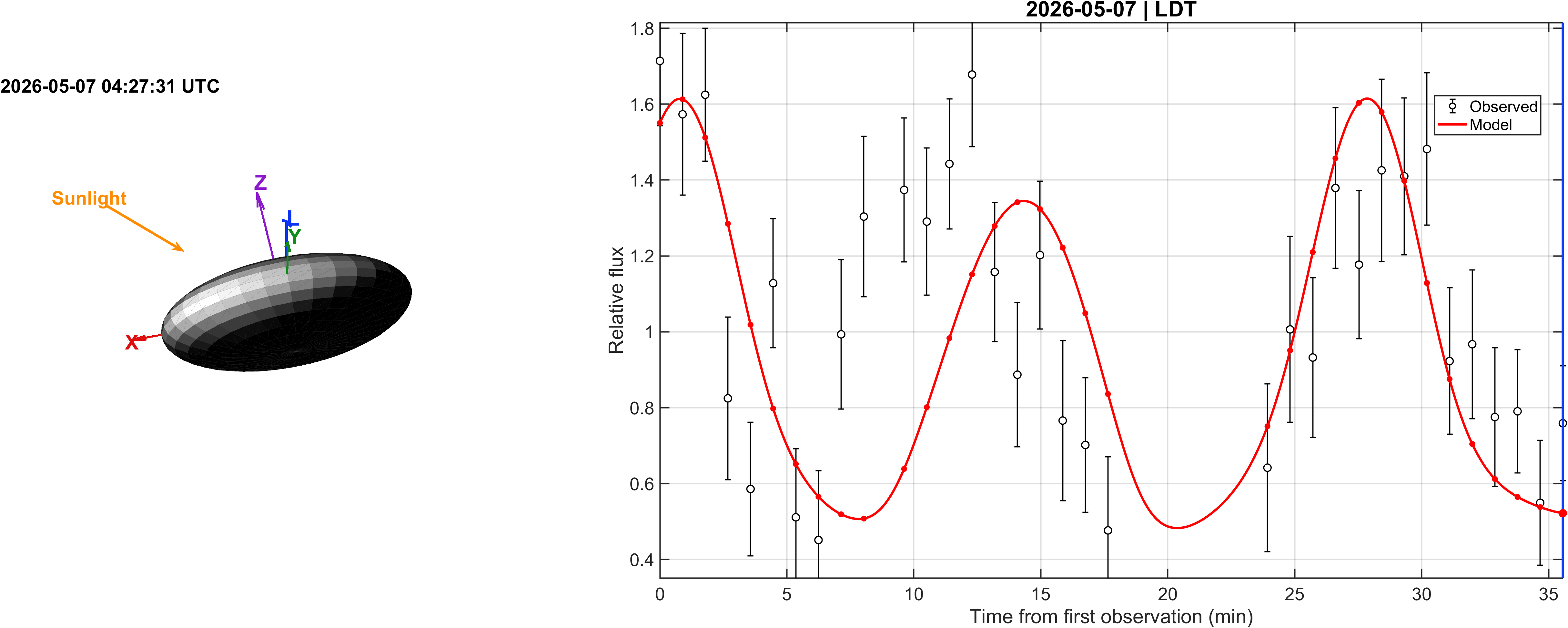}
    \caption{}
    \end{interactive}
\end{subfigure}
\caption{Animations showing the SAM rotational model and the fit to lightcurves from April 20 (a), April 22 (b), May 9 (c), and May 7 (d), respectively. The angles between the line of sight and the angular momentum vector $\mathbf{L}$ are $16^\circ$, $14^\circ$, $4.5^\circ$, and $5^\circ$ in the four epochs, respectively.  Therefore, there is a strong projection effect for the angular momentum vector, which almost directly points away from the viewer in this case.}
\label{fig:animation_sam}
\end{figure*}

We provide animations in this section to show the non-principal-axis rotation of the ellipsoidal model and the fits to lightcurves for both the LAM model solution (Fig.~\ref{fig:animation}) and SAM model solution (Fig.~\ref{fig:animation_sam}), as discussed in \S\ref{sec:rotation}.  The rotational parameters are listed in Table~\ref{tab:axis}.

\section{Convex Shape Model}
\restartappendixnumbering

\begin{figure*}[htb!]
\centering
\begin{subfigure}{0.5\textwidth}
    \begin{interactive}{animation}{20260420_GeminiN_polyhedron.mp4}
    \centering
    \includegraphics[width=0.9\linewidth]{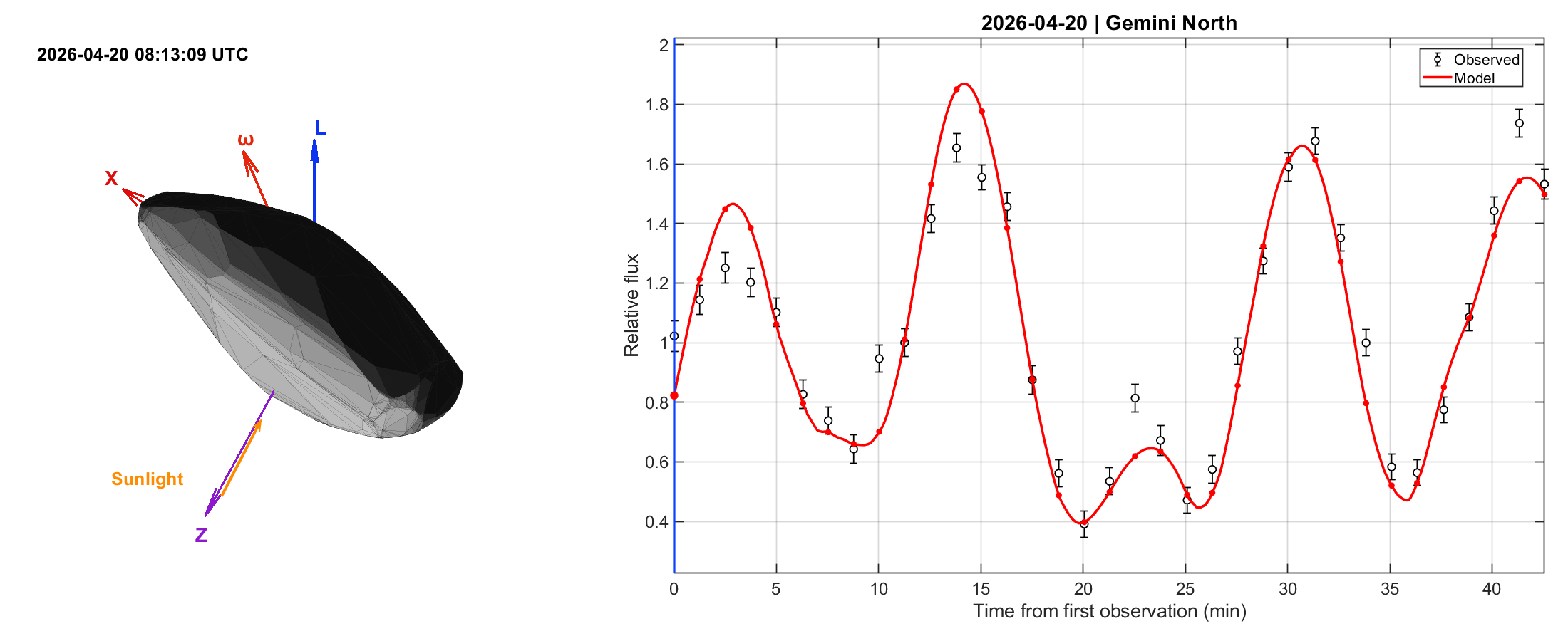}
    \caption{}
    \end{interactive}
\end{subfigure}
\hfill
\begin{subfigure}{0.5\textwidth}
    \begin{interactive}{animation}{20260422_GeminiN_polyhedron.mp4}
    \centering
    \includegraphics[width=0.9\linewidth]{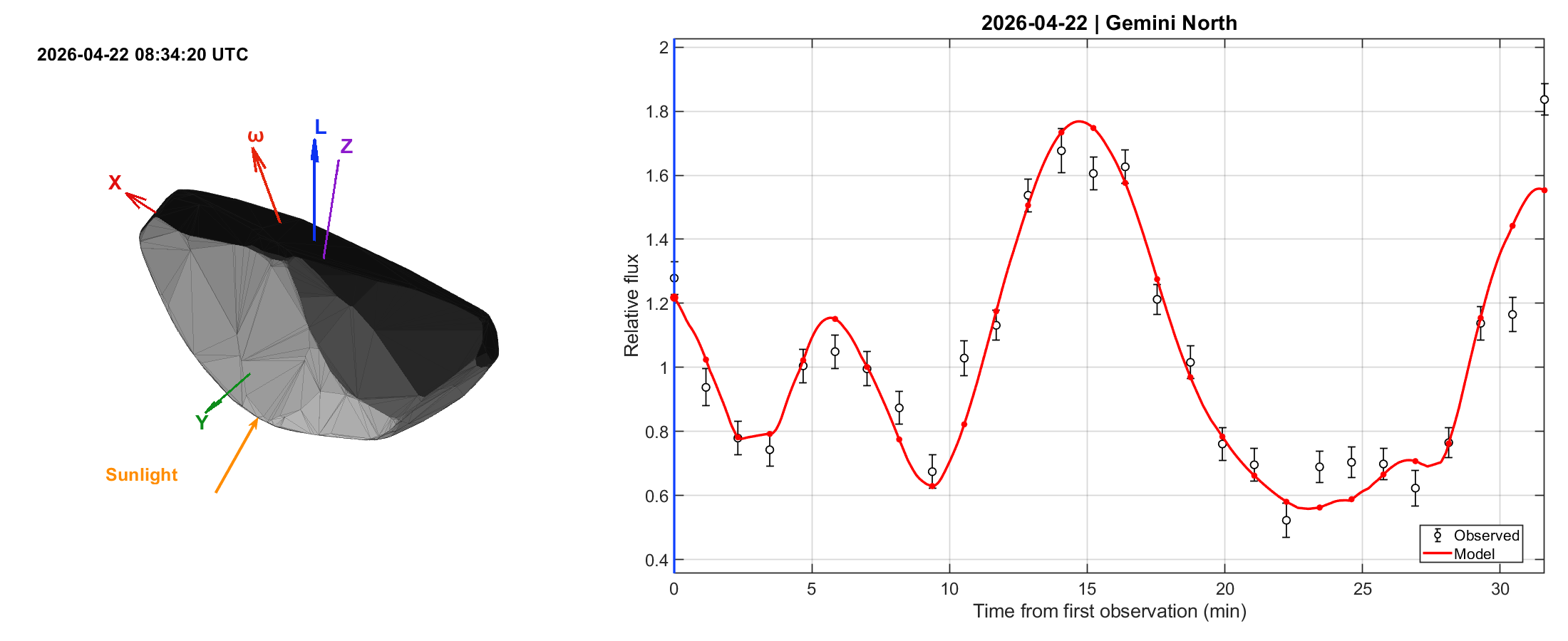}
    \caption{}
    \end{interactive}
\end{subfigure}
\vspace{0cm}
\begin{subfigure}{0.5\textwidth}
    \begin{interactive}{animation}{20260509_GeminiN_polyhedron.mp4}    
    \centering
    \includegraphics[width=0.9\linewidth]{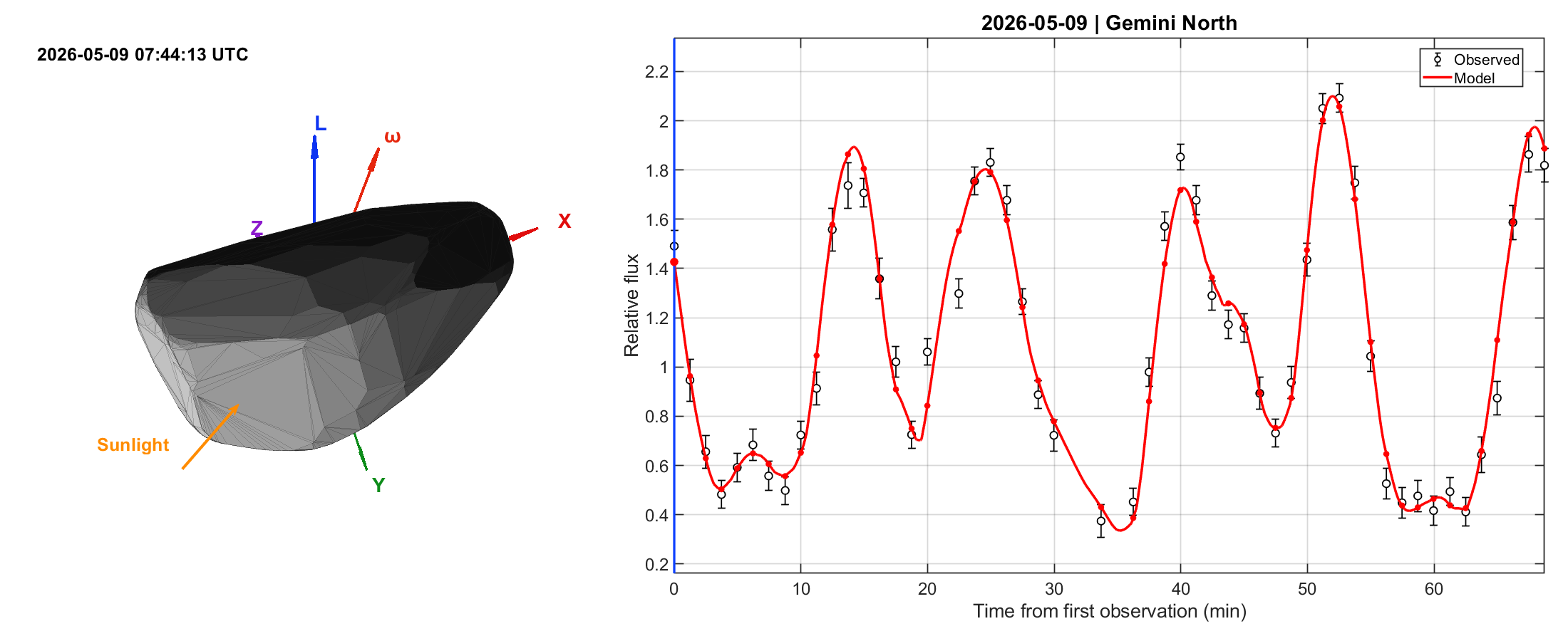}
    \caption{}
    \end{interactive}
\end{subfigure}
\hfill
\begin{subfigure}{0.5\textwidth}
    \begin{interactive}{animation}{20260507_LDT_polyhedron.mp4}
    \centering
    \includegraphics[width=0.9\linewidth]{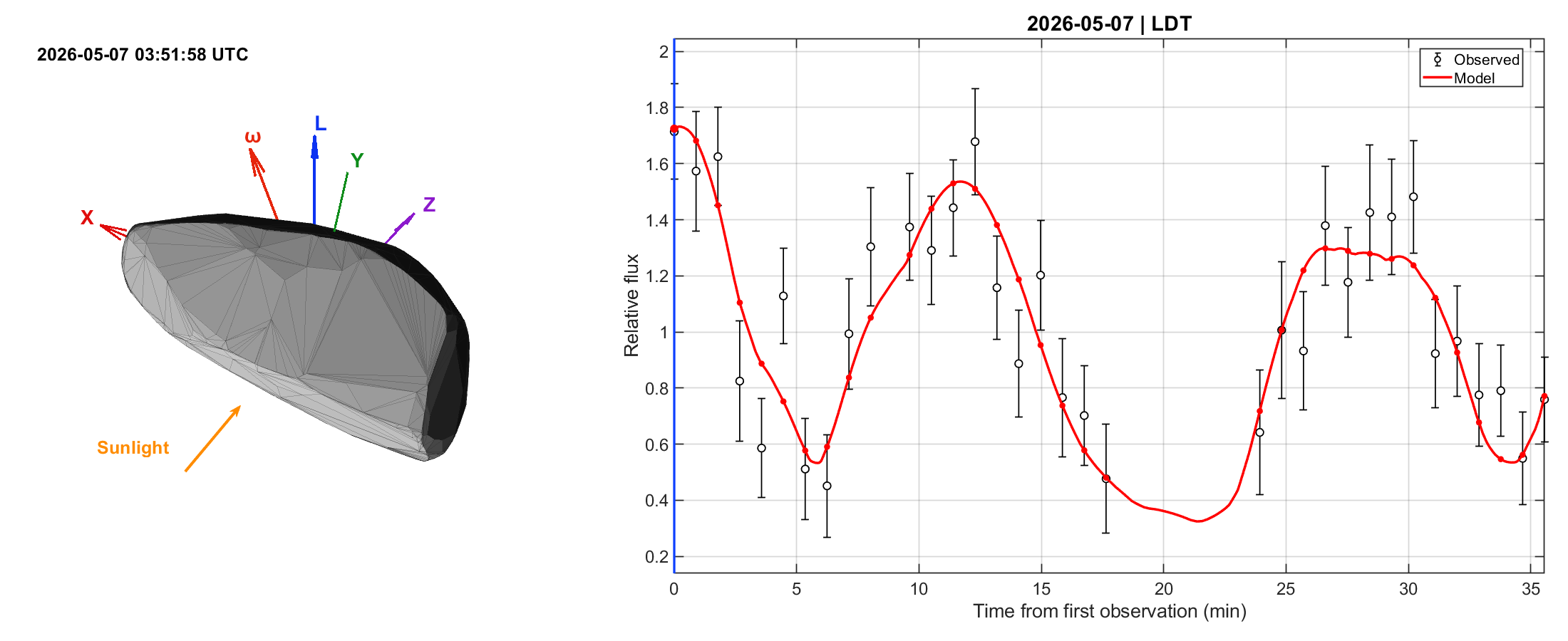}
    \caption{}
    \end{interactive}
\end{subfigure}
\caption{Animations of the rotational model with the convex shape model showing the fit to lightcurves from April 20 (a), April 22 (b), May 9 (c), and May 7 (d), respectively. Similar to the LAM model with an ellipsoidal shape, the angular momentum vector $\mathbf{L}$ almost lies within the image plane, with the angles from the line of sight $102^\circ$, $101^\circ$, $89^\circ$, and $91^\circ$ in the four epochs, respectively.}
\label{fig:animation_convex}
\end{figure*}

Same as Appendix~\ref{app:ellipsoid}, but here we show the rotational model with the convex shape model inverted from the lightcurves as discussed in \S\ref{sec:shape} (Fig.~\ref{fig:animation_convex}). This model corresponds to the solution marked as LAM as listed in Table \ref{tab:axis}, not the LAM Mirror.

\clearpage
\bibliography{refs}  
\bibliographystyle{aasjournalv7}



\end{CJK*}
\end{document}